# Awareness and Adoption of AI Technologies in the Libraries of Karnataka

Dr Felcy D'Souza
Librarian, St Joseph Engineering College, Mangalore, Karnataka
librarian@sjec.ac.in

**Abstract**

This study aims to determine the awareness and adoption of Artificial Intelligence (AI) technologies in the respondent libraries of Karnataka based on demographic variables such as gender, age, academic status, and professional experience. This study employed a survey research method to evaluate the awareness and adoption of AI technologies among the respondent library professionals in Karnataka. The study employed a stratified random sampling method to select a sample of 120 respondents from a diverse population, encompassing library professionals across multiple institution types including engineering colleges, medical colleges, and degree colleges. The Chi-square test was used to analyze the data. The study revealed that there is a statistically significant difference in the awareness and adoption of AI technologies based on the factor of gender. Whereas there no significant relationship exists between the degree of awareness and adoption of AI technologies based on factors such as age, academic ranking, and professional experience. AI-powered plagiarism detection, grammar checking, and ChatGPT are the most popularly employed AI technologies among the respondents. The respondents are of the perception that AI will support Librarians and not replace them.

## 1. Introduction

In today's fast-developing digital era, libraries play a crucial role in providing access to knowledge and information (Galibtech, 2024). By using AI technologies, libraries can give custom-made recommendations, automate daily library tasks, facilitate data-driven decision-making and thus enhance the services and functions of the libraries. The AI techniques such as natural language processing (NLP), machine learning(ML), image recognition, and data analytics are employed in the libraries. But, before applying AI technologies, libraries must consider the challenges and benefits of implementing it (Mallikarjuna, 2024).

### 1.1 Artificial Intelligence (AI) Definition

Russel and Norvig (2020) defined AI as that *"enables the machine to exhibit human intelligence, including the ability to perceive, reason, learn, and interact, etc."* Rai et al. (2019) define AI as *"the ability of a machine to perform cognitive functions that we associate with human minds, such* as perceiving, reasoning, learning, interacting with the environment, problem-solving, decision-making, and even demonstrating creativity".

## 2. Objectives of the Study

i. To study the awareness of AI technologies among the respondent college library professionals in Karnataka.

ii. To find out the adoption of AI technologies in the libraries of respondent colleges in Karnataka.

iii. To know the problems and challenges encountered by library professionals while adopting AI technologies in their libraries.

## 3. Hypotheses of the Study

i. There is no significant difference between males and females regarding the awareness and adoption of AI technologies in the respondent libraries.

ii. There is no significant relationship between age, academic ranking, and professional experience in the awareness and adoption of AI technologies in the respondent libraries.

## 4. Literature Review

Tandel (2024) discussed the usage of AI tools such as Research Rabbit, perplexity, Scite, ChatGPT, Consensus, EndNote, Semantic Scholar, Elicit, and QuillBot and found its features like citation-based mapping, free access papers, transparency reference checks, creating bibliographies, quick-locate papers, search and determine terms, grammar checks, plagiarism detectors which the librarians used to help their readers for research support. The research conducted by Poluru (2024) revealed that expert systems for reference assistants, robots designed to assist with tasks like book sorting, and the integration of virtual reality for immersive learning experiences are some of the popular AI applications in libraries. Chandrashekara and Mulimani (2024) examined the developments, benefits, problems, and prospective effects of AI on LIS services. Mali (2023) highlighted the value of openness and responsibility in the use of Chat GPT and the requirement of continual assessment and development. The study by Oseji et al. (2021) outlined the AI applications such as subject indexing, collection development, technical services, reference services, and descriptive cataloguing. The paper by Ajakaye (2021) discussed some of the features of artificial intelligence (AI), the services that libraries can offer from it, and the advantages and problems borne by the libraries when implementing it in their libraries. Barki (2022) discussed the issues of applying artificial intelligence (AI) tools like robots, and digital reality in libraries. Omame and Alex-Nmecha (2020) study contained virtual reality for more effective learning, robots that read books and shelves, and smart systems for reference. Vijaykumar and Sheshadri (2019) discussion includes speech recognition, image processing, natural language processing, robotics, expert systems, artificial neural networks, fuzzy logic, and others. Chen et al. (2021) observed that libraries incorporated with tools like the Internet of Things, RFID, Wi-Fi, BLE, natural language processing (NLP), deep learning, recommender techniques, and optical character recognition (OCR) offer innovative services. The study by Subraveerapandiyan and Gozali (2024) indicates that most LIS professionals know about AI technologies in India and opine that Artificial Intelligence can improve library activities, accessibility, and support decision-making. They believe that AI replacing human intelligence within libraries. The majority of the libraries used optical character recognition and smart shelving AI technologies. Some of the factors to be considered while applying AI in libraries are user policy, financial support, and library professionals' technology skills. The study by Jha (2023) revealed that AI is a vibrant technology that can be used in library services; but, adequate funds, librarian's attitudes, and technical skills are the obstacles that hinder the application of AI in libraries.

## 5. Research Methodology

This study adopted a survey research method to evaluate the perception of AI among library professionals in Karnataka. The study used a stratified random sampling method to select a sample population, including library professionals from engineering, medical colleges, and degree colleges. A self-designed questionnaire was shared among 200 library professionals using Google Forms as a survey instrument. The final sample comprised 120 respondents from participating libraries.

## 6. Data Analysis

**6.1 Demographic Characteristics of the Respondents**: The demographic information of the library professionals considered for the study were gender, age, educational qualification, designation, work experience, and type of organization.

**Table 1: Demographic Characteristics of the Respondents**

| Variables | | Responses | Percentage (%) |
|---|---|---|---|
| Gender | Male | 72 | 60 |
| | Female | 48 | 40 |
| Age | 21 to 30 years | 8 | 6.6 |
| | 31 to 40 years | 12 | 10.0 |
| | 41 to 50 years | 40 | 33.3 |
| | 51 years and above | 60 | 50.0 |
| Educational Qualification | Diploma in Library Science | 4 | 3.3 |
| | B.LISc | 20 | 16.7 |
| | M.LISc | 37 | 30.83 |
| | PhD | 53 | 44.17 |
| | Others degrees | 6 | 5.0 |
| Designation | Chief Librarian | 32 | 26.7 |
| | Selection Grade Librarian | 6 | 5 |
| | Librarian | 62 | 51.7 |
| | Assistant Librarian | 14 | 11.7 |
| | Library Assistant | 6 | 5.0 |
| Work Experience | Less than 5 Years | 4 | 3.3 |
| | 6 to 10 Years | 6 | 5.0 |
| | 11 to 15 Years | 22 | 18.3 |
| | 16 to 20 Years | 20 | 16.7 |
| | 21 Years to 25 Years | 30 | 25.0 |
| | More than 25 Years | 38 | 31.7 |
| Type of Organization | Government | 12 | 10.0 |
| | Private Aided | 16 | 13.3 |
| | Private unaided | 58 | 48.3 |
| | Autonomous | 22 | 18.3 |
| | Deemed to be University | 12 | 10.0 |
| **Total** | | **120** | **100.0** |

The study covered a different group of participants, with 60% males and 40% females. This inferred that male library professionals were more in terms of numbers than females (Manjunath and Patil, 2020). The respondents who were between 21 to 30 years were 8(6.6%), 31 to 40 years were 12(10%), 41 to 50 years were 40(33.3%). It is interesting to note that about 50% of the respondents are 51 years and above. Most of the respondents were qualified with PhD degrees 53(44.17%) and designated as Librarians 62(51.7%). Also, most 38(31.7%) of the respondents have more than 25 years of experience in their profession. Private unaided colleges 58(48.3%) and autonomous 22(18.3%) were the most represented type of organizations.

Out of the 120 respondents, most of the respondents had an age range between 51 years and above, which means elderly, highly qualified, and having good experience library professionals formed the greater part of the respondents of this study.

### 6.2 Awareness of AI Technologies by Respondents

A question was raised to the respondents to know their awareness of the concept of AI technologies in the libraries and the responses obtained are depicted in Table 2 below;

**Table 2: Awareness of AI Technologies by Respondents**

| No. of Respondents | | |
|---|---|---|
| **Aware** | **Not Aware** | **Total Responses** |
| 116 (97%) | 4 (3%) | 120 (100%) |

Table 2 shows that, among the 120 library professionals who responded to the questionnaire, the majority 116(97%) are aware of AI technologies in libraries.

**6.3 Awareness of the Types of AI Technologies in the Libraries**

Table 3 shows the awareness of AI technologies among the respondents as below;

Table 3: Awareness of the Types of AI Technologies in the Libraries

| Types of AI Technologies | No. of Responses & Percentage |
|---|---|
| Plagiarism checking | 94 (78.3%) |
| Grammar checking | 66 (55%) |
| ChatGpt | 62 (51.7%) |
| Text and Data Mining | 28 (23.3%) |
| Image Recognition system | 26 (21.7%) |
| Recommendation Systems | 18 (15%) |
| Knowledge Graphs | 18 (15%) |
| Smart Shelving | 16 (13.3%) |
| Speech Recognition | 14 (11.7%) |
| Chatbots | 12 (10%) |
| Natural Language Processing | 10 (8.3%) |
| None of the above | 8 (6.7%) |

The data reveal that the majority 94(78.3%) are aware of the AI-powered Plagiarism-detecting tool followed by grammar-checking 66 (55%) AI technique. The 62(51.7%) respondents are aware of ChatGPT followed by Text and data mining 28(23.3%) and image recognition systems 26(21.7%). They are aware of recommendation systems and knowledge graphs 18(15%) respectively and smart shelving 16(13.3%) and speech recognition 14(11.7%) whereas very few of them 12(10%) are aware of Chatbots and natural language processing 10(8.3%). It is observed that 8(6.7%) are not aware of any AI technologies in libraries.

**6.4 Gender-wise Impact on the Awareness and Adoption of AI Technologies among respondents.**

**Hypothesis 1:** There is no significant relationship amongst males and females regarding the awareness and adoption of AI technologies in the respondent libraries.

Table 4 shows the gender impact on the awareness and adoption of AI technologies among the respondents as below;

**Table 4: Gender-wise Awareness and Adoption of AI Technologies**

| Variable | | Status | | | Total | Chi-Square Tests | | |
|---|---|---|---|---|---|---|---|---|
| | | Aware & adopted | Aware & not adopted | Not Aware & not adopted | | Value | df | p-value |
| Gender | Male | 66 (91.7%) | 6 (8.33%) | 0 | 72 (100%) | 6.226 | 2 | .044 |
| | Female | 40 (83.34%) | 4 (8.33%) | 4 (8.33%) | 48 (100%) | | | |

It is depicted from Table 4 that, 66(91.7%) of the male library professionals are aware of and adopted the AI technologies in their libraries against 40(83.34%) of the females. At the same time, 8.33% of the male and female respondents are aware of AI technologies but have not adopted them due to various reasons. The p-value is 0.044, which is less than the conventional threshold of 0.05. This indicates that there is a statistically significant association between gender and the awareness and adoption status categories. Hence hypothesis 1 is rejected. Based on the chi-square test, we can conclude that gender has a significant impact on the awareness and adoption status of the individuals.

**6.5 Age-wise, academic ranking, and professional experience-wise impact on the awareness and Adoption of AI Technologies among respondent Libraries**

**Hypothesis 2:** There is no significant relationship between age, academic ranking, and professional experience in the awareness and adoption of AI technologies in the respondent libraries.

Table 5 shows the Age-wise, academic ranking, and professional experience-wise impact on the awareness and adoption of AI technologies among the respondents as below;

**Table 5: Age-wise, academic ranking and professional experience-wise impact on the awareness and Adoption of AI Technologies**

| Variable | | Status | | | Total | Chi-Square Tests | | |
|---|---|---|---|---|---|---|---|---|
| | | Aware & adopted | Aware & not adopted | Not Aware & not adopted | | Value | df | p-value |
| Age in Years | 21 to 30 | 6 (75%) | 2 (25%) | 0 | 8 (100%) | 5.415 | 6 | .492 |
| | 31 to 40 | 12 (100%) | 0 | 0 | 12 (100%) | | | |
| | 41 to 50 | 34 (85%) | 4 (10%) | 2 (5%) | 40 (100%) | | | |
| | 51 & above | 54 (90%) | 4 (6.67%) | 2 (3.33%) | 60 (100%) | | | |
| | Chief Librarian | 28 | 2 | 0 | 30 | 7.114 | 8 | .524 |
| Designation | Selection Grade Librarian | 6 | 2 | 0 | 8 | | | |
| | Librarian | 60 | 6 | 4 | 70 | | | |
| | Assistant Librarian | 6 | 0 | 0 | 6 | | | |
| | Library Assistant | 6 | 0 | 0 | 6 | | | |
| Work Experience | Less than 5 Years | 4 | 0 | 0 | 4 | 15.720 | 10 | .108 |
| | 6 to 10 Years | 4 | 2 | 0 | 6 | | | |
| | 11 to 15 Years | 20 | 0 | 2 | 22 | | | |
| | 16 to 20 Years | 16 | 4 | 0 | 20 | | | |
| | 21 to 25 Years | 28 | 2 | 0 | 30 | | | |
| | More than 25 Years | 34 | 2 | 2 | 38 | | | |

It is observed from Table 5 above that the p-values for age, academic ranking, and professional experience are all greater than 0.05. Therefore, the null hypothesis is accepted, which states no significant relationship exists between age, academic ranking, and professional experience in the awareness and adoption of AI technologies in the respondent category of libraries in Karnataka.

**6.6 Library Professionals Opinions on the Application of AI Technologies in Libraries**

Table 6 depicts the views of the respondents on the implementation of AI technologies in their libraries below;

**Table 6: Library Professionals Opinions on the Application of AI Technologies in Libraries**

| Library Professionals Opinions | No. of Responses & Percentage |
|---|---|
| AI enhancing library activities and services | 86 (71.7%) |
| AI supporting Librarians | 80 (66.7%) |
| AI will enhance accessibility to resources | 66 (55%) |
| AI will enhance academic/research activities | 66 (55%) |
| Willingness to learn more about AI tools | 54 (45%) |
| AI will not replace Librarians | 22 (18.3%) |

The respondents opine that the application of AI techniques will enhance library activities and services 86 (71.7%). It is also found that AI technologies will enhance accessibility to library resources and improve the academic/research activities 66(55%) of the Institute respectively. It is also observed that 54(45%) of the respondents have a willingness to learn more about AI tools. The respondents are of the perception that AI will support Librarians 80(66.7%) and not replace them 22 (18.3%).

**6.7 Problems and Challenges for Application of AI Technologies in Libraries**

Table 7 describes the factors affecting the implementation of AI technologies in libraries as below;

**Table 7: Problems and Challenges for Adoption of AI Technologies in Libraries**

| Problems and Challenges | Yes (%) |
|---|---|
| Skilled Library Professionals | 76 (63.3%) |
| Financial Constraints | 62 (51.7%) |
| Librarians willingness to adopt AI tools in libraries | 40 (33.3%) |
| User Policy | 36 (30%) |
| Support from Management | 32 (26.7%) |

It is presumed that some necessary factors must be considered when employing AIs in libraries (Yoon et al., 2021). The respondents expressed that (Table 7), skilled library professionals 76(63.3%), financial constraints 62(51.7%), librarian willingness to adopt AI technologies in libraries 40(33.3%), user policy 36(30%) and support from management 32(26.7%) are the problems and challenges faced by the library professionals for adopting AI technologies in the respondent libraries.

**6. Findings of the Study**

i. The male respondents were the biggest participants of this study than females.

ii. Most of the respondents are 51 years and above, highly qualified, and have a professional experience of more than 25 years and above. This means elderly, highly qualified, and having good professional experience formed the greater part of the respondents of this study.

iii. A majority of 97% of the respondents were aware of the AI technologies in libraries.

iv. Plagiarism-detecting, Grammar checking, and ChatGPT are more widely incorporated AI-powered technologies in the respondent libraries.

v. It is revealed from the study that gender has a significant impact on the awareness and adoption of AI technologies in libraries. Hence hypothesis 1 is rejected.

vi. It is observed that the p-values for age, academic ranking, and professional experience are all greater than 0.05. Therefore, null hypothesis 2 is accepted, which states no significant relationship exists between age, academic ranking, and professional experience in the awareness and adoption of AI technologies in the respondent category of libraries in Karnataka.

vii. The respondents are of the perception that AI will support Librarians 80(66.7%) and not replace them 22 (18.3%).

viii. Skilled library professionals 76(63.3%), financial constraints 62(51.7%), librarian willingness to adopt AI technologies in libraries 40(33.3%), user policy 36(30%) and support from management 32(26.7%) are some of the problems and challenges faced by the library professionals for adopting AI technologies in their libraries.

## 7. Conclusion

The study recommends that library professionals require training on AI tools and technologies for this they should have support from their Universities or Institutes to organize some seminars, workshops, and skill development programmes for the library professionals. Integrating AI-powered technologies will help libraries optimize library resources, offer innovative library services, and enhance research in the Institute. The study suggests collaboration of library professionals with academicians, researchers, and policymakers before implementing AI technologies in libraries.